\documentclass[preprint,preprintnumbers,amsmath,amssymb]{revtex4}

\usepackage{graphicx}
\usepackage{amssymb}
\usepackage{inputenc}
\usepackage{hyperref}

\begin{document}

\title{A GEANT4 based Simulation for Pixelated \\ X-ray Hybrid Detectors: Extended Report}

\author{F. Marinho}
\email{marinho@macae.ufrj.br}
\address{Universidade Federal do Rio de Janeiro, Av. Aluizio Gomes, 50, 27930-560, Maca\'e, RJ, Brazil}
\author{K. Akiba}
\address{Universidade Federal do Rio de Janeiro, Av. Athos da Silveira Ramos, 149, 21941-972, Rio de Janeiro, RJ, Brazil}

\begin{abstract}

Monte Carlo simulations are powerful tools for understanding the effects of radiation interactions within detector devices allowing not only to evaluate typical estimates for experimental measurements and to serve as means for designing experiments but to provide additional information that are usually not easily accessible experimentally. A simple numerical approach to describe the functioning of pixelated detectors is presented in this paper. Estimates from a variety of simulated setups are obtained and the observed features compared with experimental results to verify to which extent they are correctly described by this method. Fundamental curves such as interaction probability and reconstructed energy deposited are calculated as function of beam energy. Pixel response, charge sharing effects and resolution estimates are also obtained.

\bigskip
{\bf keywords:} Pixel detector, X-ray radiation, Medipix, Monte Carlo simulation
\end{abstract}

\maketitle

\section{Introduction}\label{Intro}

The recent development of processing architectures and of software for description of radiation-matter interactions made the implementation of simulation algorithms very attractive giving the short running periods of time on a single computer, local cluster or world wide grids. A Monte Carlo simulation based on the Geant4 package \citep{geant4} was developed for emulating the functioning of energy sensitive x-ray hybrid pixel detectors. The approach adopted allows to reproduce detailed features of the radiation interactions inside the sensor matrix and the detector response takes into account electronics readout effects. These detectors are used on a broad range of applications since they can provide adequate energy and position resolution with fast data acquisition, therefore such computational endeavors are of paramount importance as tools for detector research activities possibly indicating optimized parameters for a given technology application such as sensor material, width, pixel dimension, etc. 

As an example of this approach we simulated a $\rm 300 \mu m$ silicon sensor bump bonded to a medipix/timepix family chip \citep{medipix, medipix2}. The Geant4 package was used to describe the propagation of particles within the different materials. The detector electronics readout was simulated with a coupled algorithm which reproduces its effects on the final energy resolution, charge sharing and related features \citep{medipix23, jjakubek}. 

A complete description of the detector in the simulation is given in section \ref{simulation} where details of code implementation are also presented. The expected energy per pixel spectrum, interaction probability and energy reconstruction as a function of the incoming energy are presented in section \ref{experiment}. Results for the simulation of a few setups were also obtained mainly related to studies for general characterization of the device and also to its performance on position resolution. Final considerations, conclusions and perspectives are discussed in section \ref{conclusion}.

\section{Simulation Description}\label{simulation}

The typical approach to implement a simulation for the passage of radiation through matter consists on the description of the materials geometry and chemical composition and appropriate particle interaction modeling including all possible scattering processes. In addition, the simulation of a detector must take into account the data acquisition which includes effects occurring on charge collection, signal amplification and readout from the electronics.

\subsection{Geometry}

The hybrid device taken as study object was a silicon 300 $\rm \mu m$ width detector with $\rm ~256 \times ~256$ pixels of 55 $\rm \mu m$ size with its surface covered by a $\rm 4 \mu m$ thin aluminum layer and its back bump bonded to a $\sim$ 1mm thick readout electronics simply represented by a rectangular block. All the materials compositions are included with their correspondent density in this description.

\subsection{Interactions}

The electromagnetic interactions were simulated using the Penelope \citep{penelope} package implementation available in Geant4. Particles propagation was performed until a final energy cut of about 100 eV. At this energy value electrons free mean path are such that the initial size of the produced charge carriers clouds are properly simulated before they are dragged towards the collection electrodes. Table 1 describes the possible processes experienced by photons and charge carriers in the detector material.

\begin{table}[!htb]
\caption{Expected electromagnetic processes which can occur within the materials.}
\label{table1} 
\centering
\begin{tabular}{c||c} \hline \hline
Particle       & Scattering \\ \hline
Photon           & Photo-electric, Compton, Conversion, Rayleigh      \\
Electron         & Ionization, Bremsstrahlung                   \\\hline\hline
\end{tabular}
\end{table}

Diffusion and charge carriers transport are performed after the aforementioned threshold is achieved and is described by equation \ref{diffusion} when the semiconductor is fully depleted:

\begin{equation}\label{diffusion}
\Delta(z)=\frac{KTd^2}{ev_d}log\left(\frac{v_d+v_{bias}}{v_d+v_{bias}-2\frac{v_dz}{d}}\right),
\end{equation}

\noindent{where $K$ is the Boltzmann constant, $T$ is the material temperature, $d$ is the sensor thickness, $v_d$ the depletion voltage, $v_{bias}$ is the applied voltage in the electrodes and $z$ is the interaction depth. Similar procedure can be performed if the device is just partially depleted.}

\subsection{Data Acquisition}\label{daq}

Another important aspect of the simulation is the detector electronics response which is observed through the charge collection in the electrodes of the detector. Precise information about the location of the interaction region and deposited energy in the device is obtained through these measurements.

A few thousands of charge carriers are produced when a particle interacts depositing part of its energy in the detector material. This amount is proportional to the deposited energy and for a silicon detector the electron-hole energy production is of $\sim \rm 3.6 ~eV$. Charge carriers recombination is avoided as the electric field in the semiconductor bulk drives them towards the opposite electrodes coupled to the sensor. Of course, the total charge can be distributed for more than one pixel depending on the interaction position of the particle within material and impinging energy. In this approach carriers are geometrically divided among pixels according to a least distance criteria with respect to the pixels center position. Adequate pixel clustering allows precise position and deposited energy reconstruction. Electronic noise and fine-tuned threshold dispersion are considered in the simulation. Multiple interactions from the primary photons are also taken into account.

The reconstructed information from the pixels obtained from the simulation are stored. The so called Monte Carlo truth is also kept and includes the correct hit position, deposited energy and type of interaction occurred. Therefore, expected theoretical quantities and the reconstructed estimates obtained via simulation can be compared and even resolutions and efficiencies can be calculated for the different operation modes of the device. This way additional information, which is not available through experimental data analysis, can be accessed allowing a more complete interpretation of the estimates obtained.

\section{Simulation of Experimental Setups}\label{experiment}

A set of experiments were simulated employing this approach to describe the pixel detector in order to evaluate its feasibility. The following sections present the description of each setup and the results obtained from the simulation. Comparison between the main features obtained and results available in the literature is provided together with additional information delivered from the simulation.

\subsection{Monochromatic Beam}

\begin{figure}[!htb]
\begin{center}
{\includegraphics[width=0.45\textwidth]{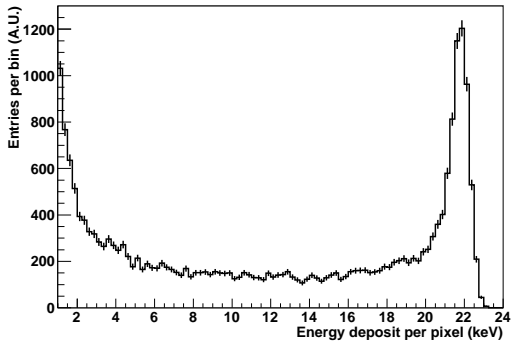}}
\end{center}
\caption{Reconstructed energy spectrum per pixel.} 
\label{enerdist}
\end{figure}

The first test for the simulation was to simply bombard the whole surface of the sensor with a monochromatic x-ray beam. Figure \ref{enerdist} shows the spectrum of the energy response per pixel. Note that almost all of the photon energy is deposited in a single pixel but a small part can be collected by a neighbor pixel such that a peak at the very low edge of the spectrum complements the main peak which is centered at a value slightly below the original beam energy (22 keV). For this detector family threshold values can be adjusted to typical values of a few keV where noise free acquisition still is possible. Contributions from three and four pixel clusters are also present but less obvious to visually spot on the spectrum. The shape of the energy spectrum obtained is very similar to those observed in \citep{ltlustos, tmichel}. The electronics smearing adopted for the single pixel energy resolution was $\rm\sim$ 100e-. Charge sharing also contributes to the broadening of the spectrum around the energy peak region. 

\begin{figure}[!htb]
\begin{center}
{\includegraphics[width=0.45\textwidth]{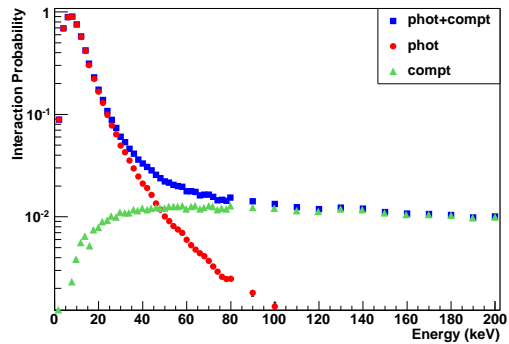}}
\end{center}
\caption{Interaction probability as a function of the incoming radiation energy (squares). Photoelectric (circles) effect and Compton scattering (triangles) contributions are shown separately.} 
\label{probinter}
\end{figure}

\begin{figure}[!htb]
\begin{center}
{\includegraphics[width=0.45\textwidth]{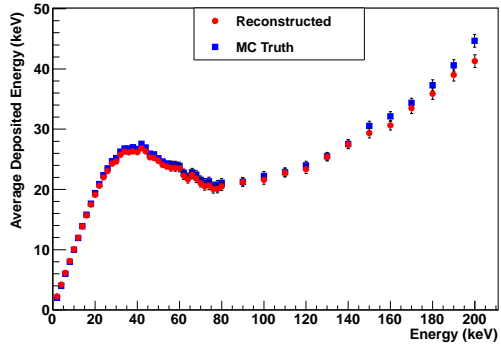}}
\end{center}
\caption{Total deposited energy as a function of the incoming radiation energy. Reconstructed (circles) and Monte Carlo truth (squares) values are shown.} 
\label{enerdep}
\end{figure}

Two other quantities often used in the design of experiments are the interaction probability and the average energy deposited in the detector material as a function of the incoming beam energy. Figure \ref{probinter} shows the total interaction probability in the detector bulk (squares) and its separated components due to the photoelectric effect (circles) and Compton scattering (triangles). Note that the Monte Carlo truth information allows to separate the full interaction probability in these two components. The photoelectric effect is the dominant process up to 50 keV while the Compton scattering contribution rises and becomes steady above $\sim$ 30 keV. Figure \ref{enerdep} shows the average energy deposited in the silicon sensor as a function of beam energy where the squares illustrate the estimate obtained for the actual values calculated from the theoretical models used. The circles represent the reconstructed energy curve obtained from the full simulation including all effects of the data acquisition mentioned in section \ref{daq} given in energy units converted from the total charge collected (1e- $\rm \sim$ 3.6eV). Across the whole energy range adopted it is verified exceptional linearity of the detector response as the two curves are almost the same. Below $\sim \rm 25 ~keV$ the deposited energy is equivalent to the energy of the incoming particle as the photoelectric effect dominates and the slope for the curve is 1. Above this value the Compton scattering contribution smoothly increases while the photoelectric effect components reduces abruptly. This behavior causes a reduction of the average energy within the $\rm 35 ~keV \le E_{beam} \le 130 keV$ range for the beam energy as the Compton scattering energy deposition is always below the incoming particle energy. However, the average energy deposit due to the Compton scattering increases with energy as this behavior can be seen in the range above $\rm 90 ~keV$.

The curves shown in Figures \ref{probinter} and \ref{enerdep} are very similar to the ones obtained in the experiments presented in \citep{mfierdele}. Moreover, information related to the types of interactions involved and reconstruction effects can be evaluated and used to accurate assert the device features and also to provide a more complete characterization of future measurements.

\subsection{Low Angle Incidence}

An oblique beam of particles impinging on the detector surface can offer interesting means to study few geometrical aspects related to the sensor characteristics and to check for the homogeneity of the sensor \citep{jjakubek}. Estimates of cluster size dependency with interaction depth, beam energy and bias voltage are accessible by this technique using a narrow beam.

\begin{figure}[!htb]
\begin{center}
{\includegraphics[width=0.32\textwidth]{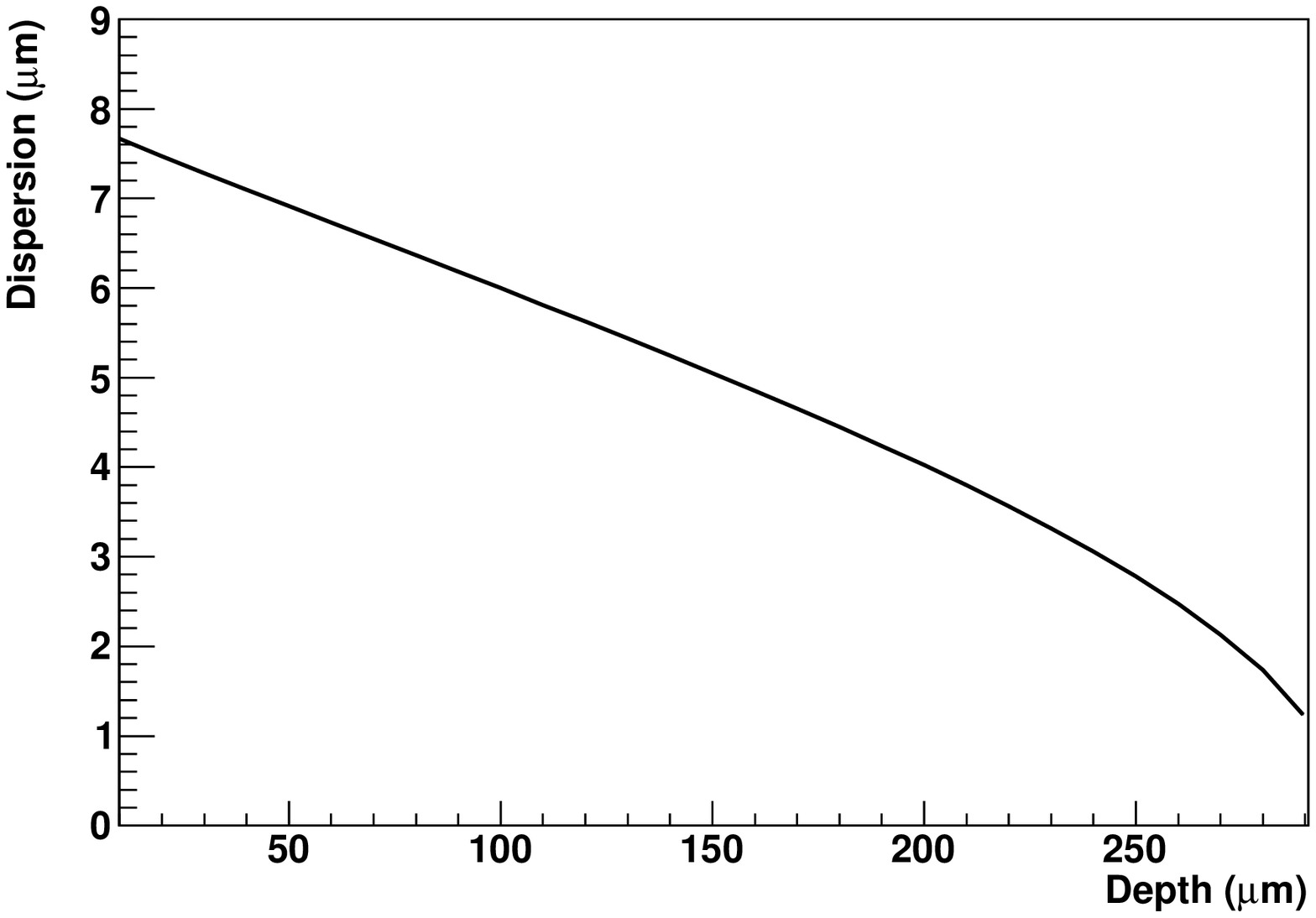}
\includegraphics[width=0.32\textwidth]{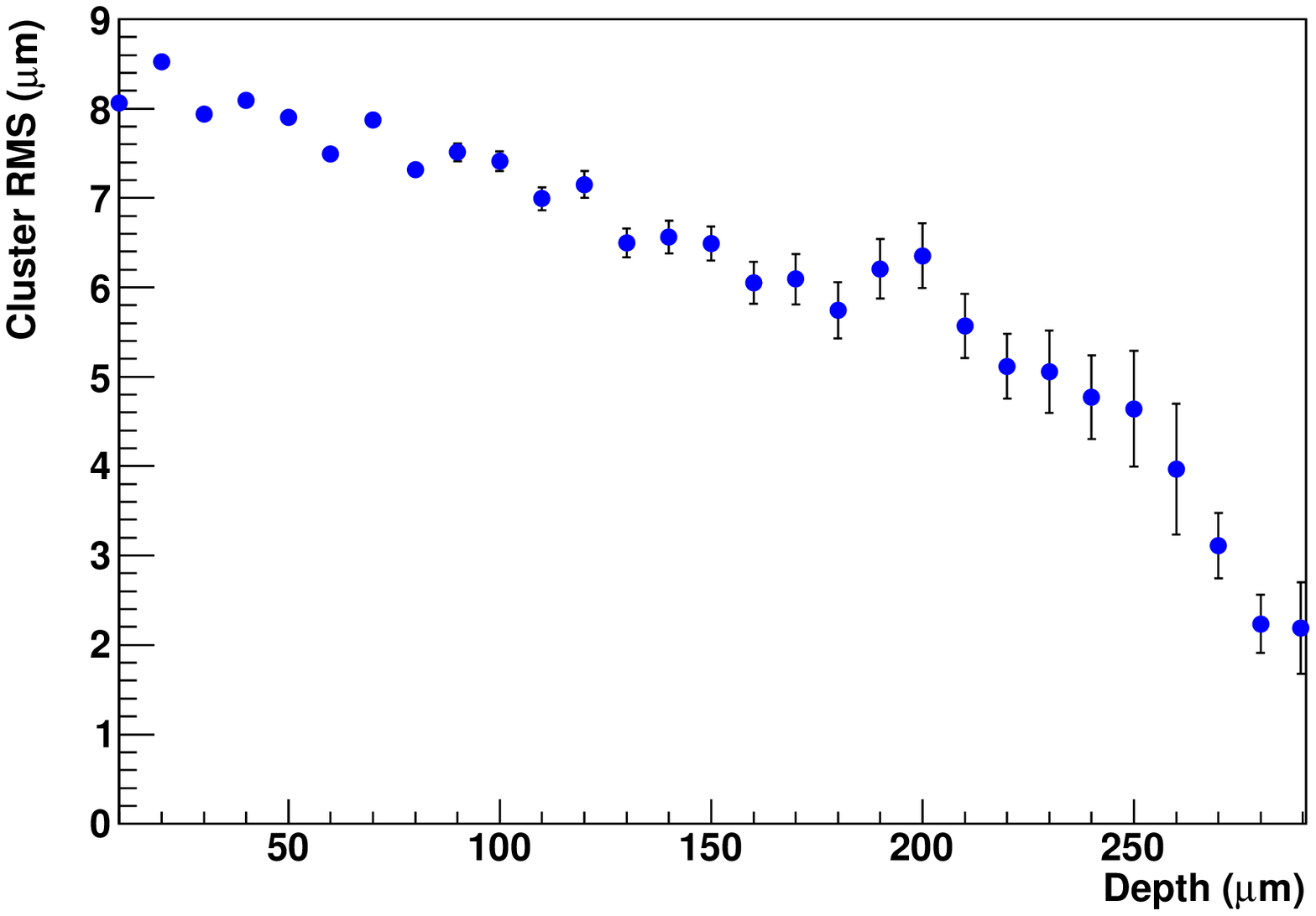}
\includegraphics[width=0.32\textwidth]{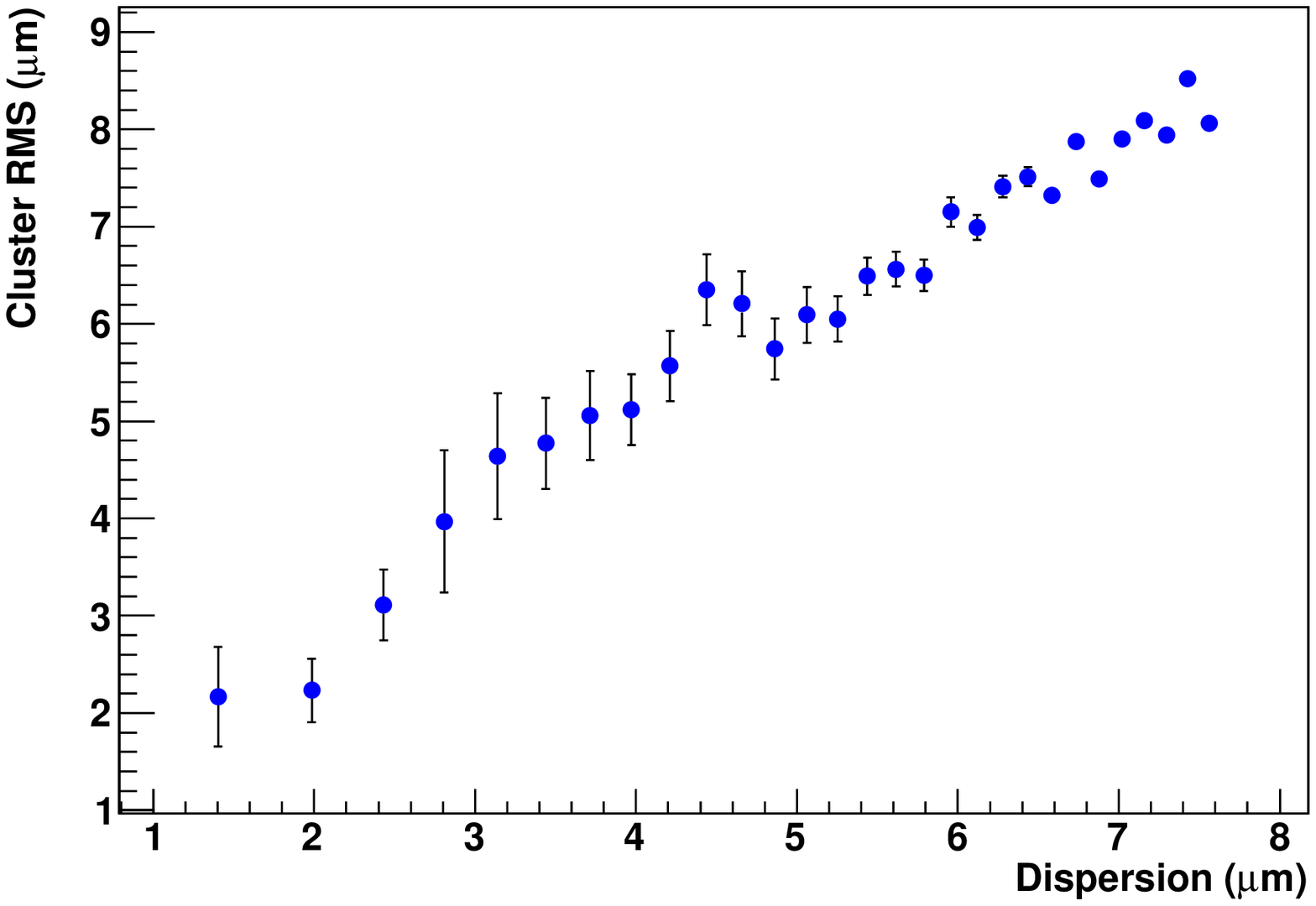}}
\end{center}
\caption{Dispersion per hit (left) and average cluster size (center) as functions of $z$. Correlation between cluster size and dispersion (right).} 
\label{clusterdep}
\end{figure}

Figure \ref{clusterdep} shows the relationship between the average reconstructed cluster size with the transverse dispersion of the charge carriers cloud suffered across its path towards the collection terminals. The expected dispersion calculated with Equation \ref{diffusion} as a function of the interaction depth $z$ is shown on the left hand side. The reconstructed cluster size is shown at the center and was calculated by taking the RMS of the charge collected per cluster in each bin. Note the almost linear correlation seen on the right hand graph.

Similar correlations can be obtained by varying the beam energy but the overall curve would be just slightly displaced in the cluster size axis as the size of the initial charge carrier cloud depends on the impinging energy. Note that the dependence with the applied bias voltage is already included in Equation \ref{diffusion} together with sensor material temperature.

\begin{figure}[!htb]
\begin{center}
{\includegraphics[width=0.45\textwidth]{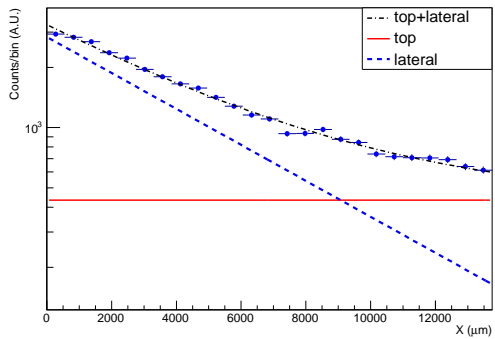}}
\end{center}
\caption{Number of hits as function of the position in the x-axis direction of the sensor.} 
\label{cntdepth}
\end{figure}

Very low angle incidence provides another interesting setup that can be useful to monitor the attenuation coefficient in different regions of the sensor as one can take advantage of the pixel structure of the device \citep{mfierdele}. The idea consists of using a wide beam to hit the device with a small angle with respect to the plane of the sensor. Figure \ref{cntdepth} shows the number of hits as a function of the distance traveled by the radiation inside the detector material. Equation \ref{counts} is used to model the distribution of hits for an homogeneous material.

\begin{equation}\label{counts}
N(x) = A + B {~\rm exp}^{-Cx}
\end{equation}

\begin{figure}[!htb]
\begin{center}
{\includegraphics[width=0.45\textwidth]{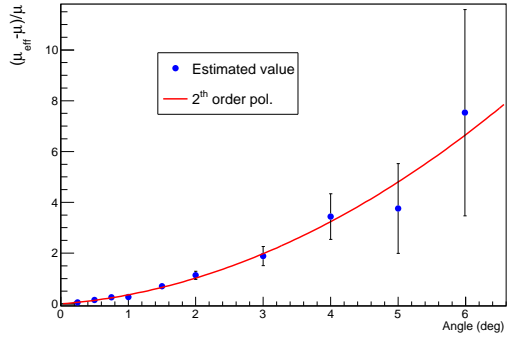}}
\end{center}
\caption{Effective attenuation coefficient as function of the radiation incident angle.} 
\label{effmu}
\end{figure}

{\noindent where $A$ represents the flux contribution impinging on the top surface of the detector and the exponential component is due to the lateral incidence in the material with $C$ being the attenuation coefficient. The fitted $B$ value is represented by the horizontal solid line while the exponential term is represented by the dashed line. Simulation shows that to obtain accurate estimates for $C$ the angle should be smaller than $\rm 0.5^o$ such that any angular dependence for this estimator can be neglected. However, the simulation shows a second order polynomial angular dependence up to $\rm 6^o$ as seen in Figure \ref{effmu}.}

Similar procedures can be employed to study defects on high-Z sensor materials \citep{efrojdh, efrojdh2}  and also to monitor the detector characteristics as it is irradiated along time and its material properties can change according to a possible non-homogeneous radiation dose accumulated at different regions of the detector geometry \citep{velo}. However, there are not such measurements for thin silicon pixel hybrids available for comparison in the literature yet.

\subsection{Pixel Sensitivity and Charge Sharing}\label{pixelsens}

Geometrical mapping of the pixel sensitivity can be obtained by shinning a narrow beam perpendicular to the detector, scanning within its area and vicinities. Such procedure can also provide information to better understand charge sharing by looking at the total count of hits in the detector \citep{macraighne, egimenez}.

\begin{figure}[!htb]
\begin{center}
{\includegraphics[width=0.45\textwidth]{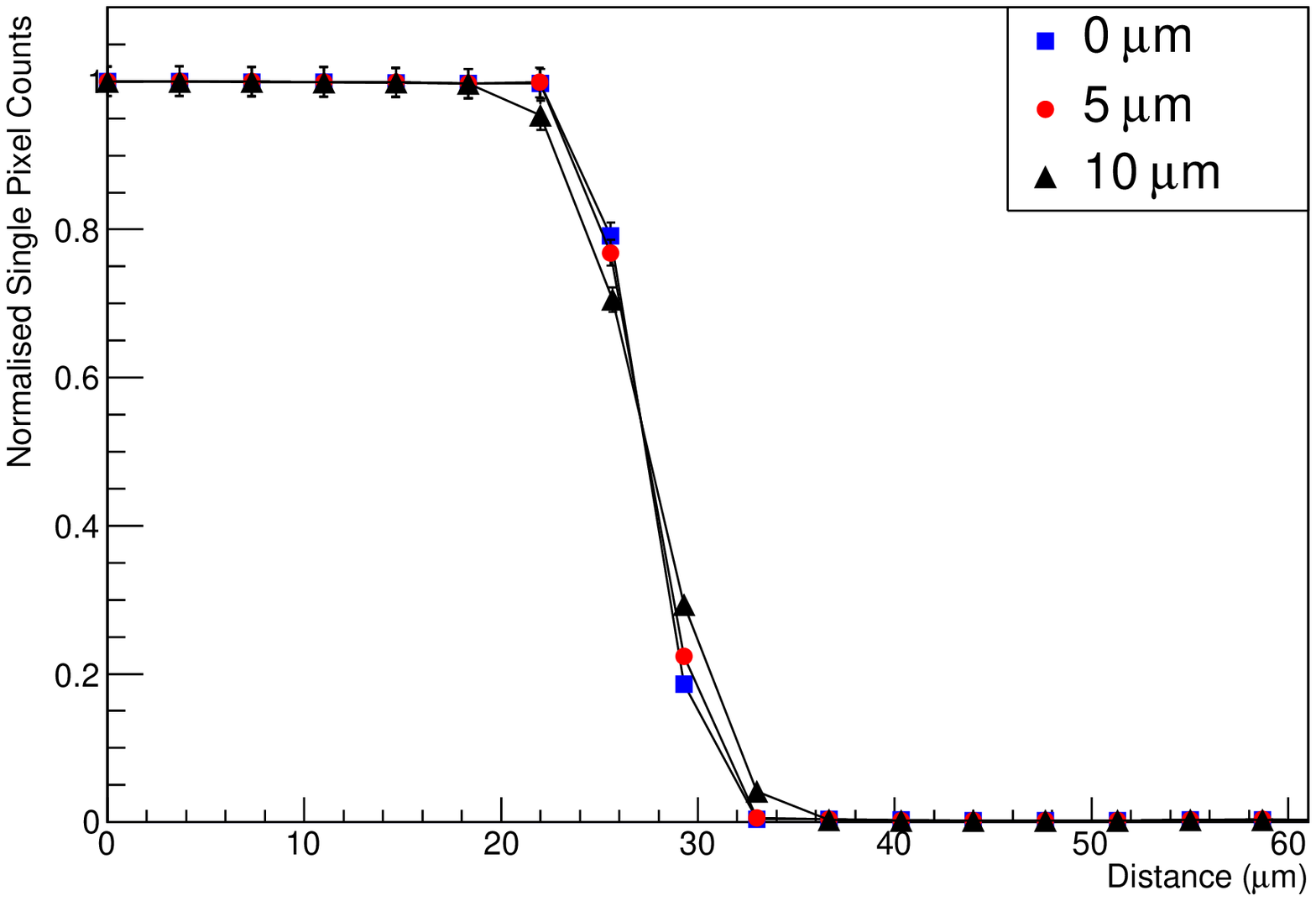}
\includegraphics[width=0.45\textwidth]{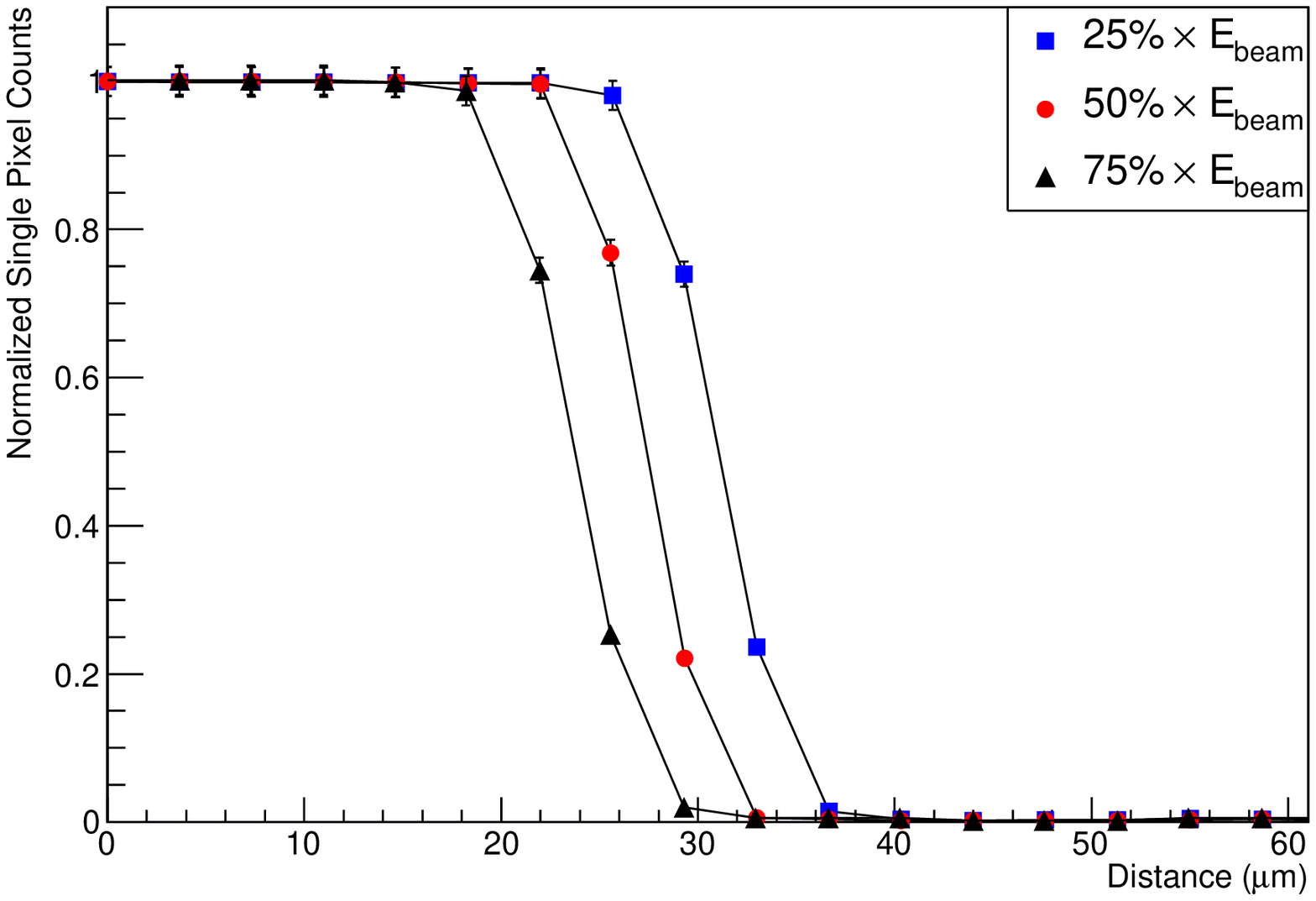}}
\end{center}
\caption{Single pixel sensitivity as function of beam position. Dependence with beam size (left) and with applied threshold (right) is illustrated.} 
\label{sensitivity}
\end{figure}

Figure \ref{sensitivity} shows a single pixel response as function of the position of beam incidence with respect to the center of the pixel. The dependence of the edge sharpness with the beam width and threshold value is illustrated on the left hand side. The beam width sizes simulated were 0, 5, and 10 $\rm \mu m$. The midpoint between the two pixels is at 27.5 $\rm \mu m$ where the neighbor pixel is also hit. On the right hand side it is seen that the loss of efficiency at the edges of the pixel, as observed in \citep{egimenez} and reproduced by the simulation, depends on the adopted threshold value with respect to beam energy. The threshold values were set to 25\%, 50\% and 75\%.  

\begin{figure}[!htb]
\begin{center}
{\includegraphics[width=0.32\textwidth]{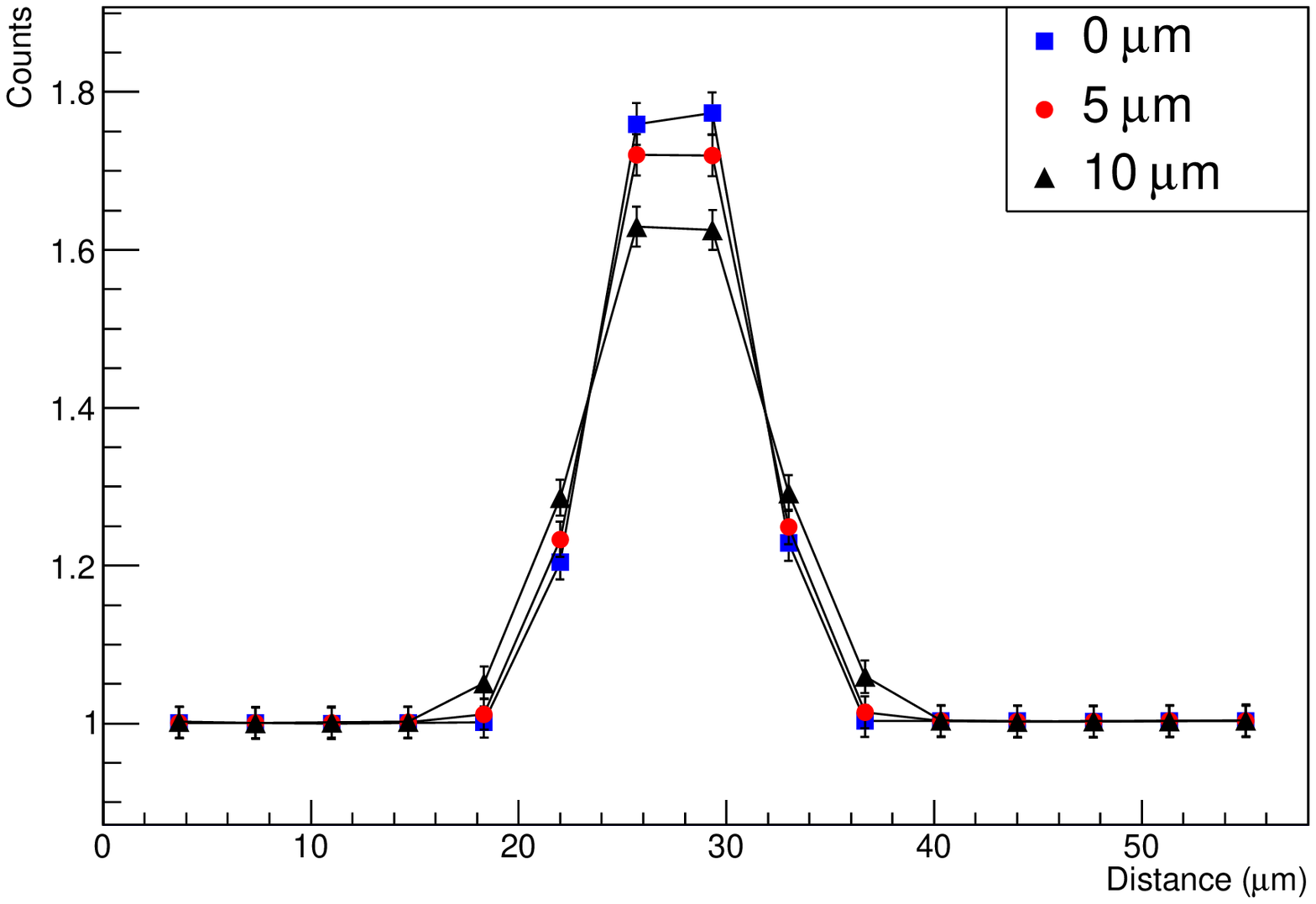}
\includegraphics[width=0.32\textwidth]{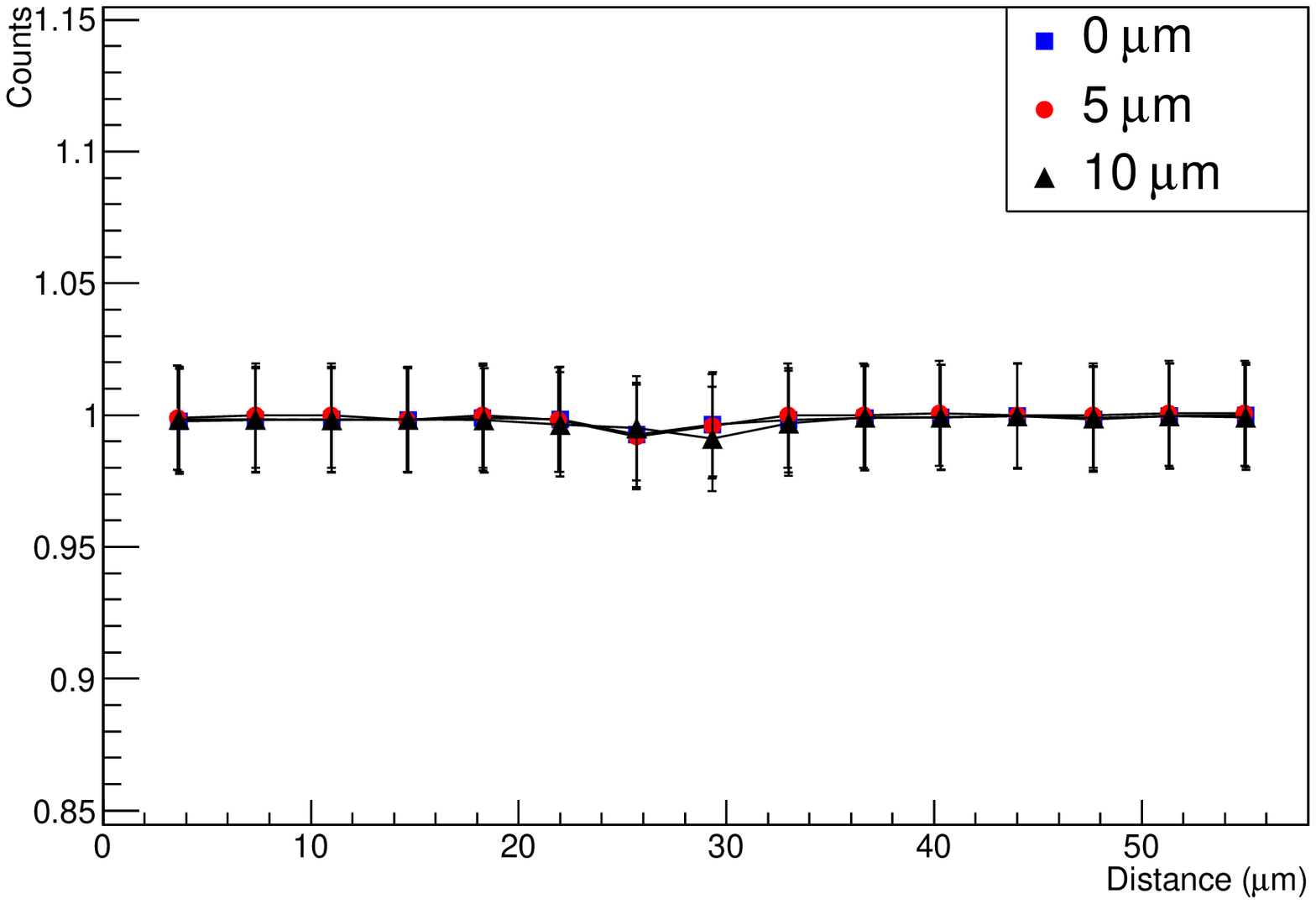}
\includegraphics[width=0.32\textwidth]{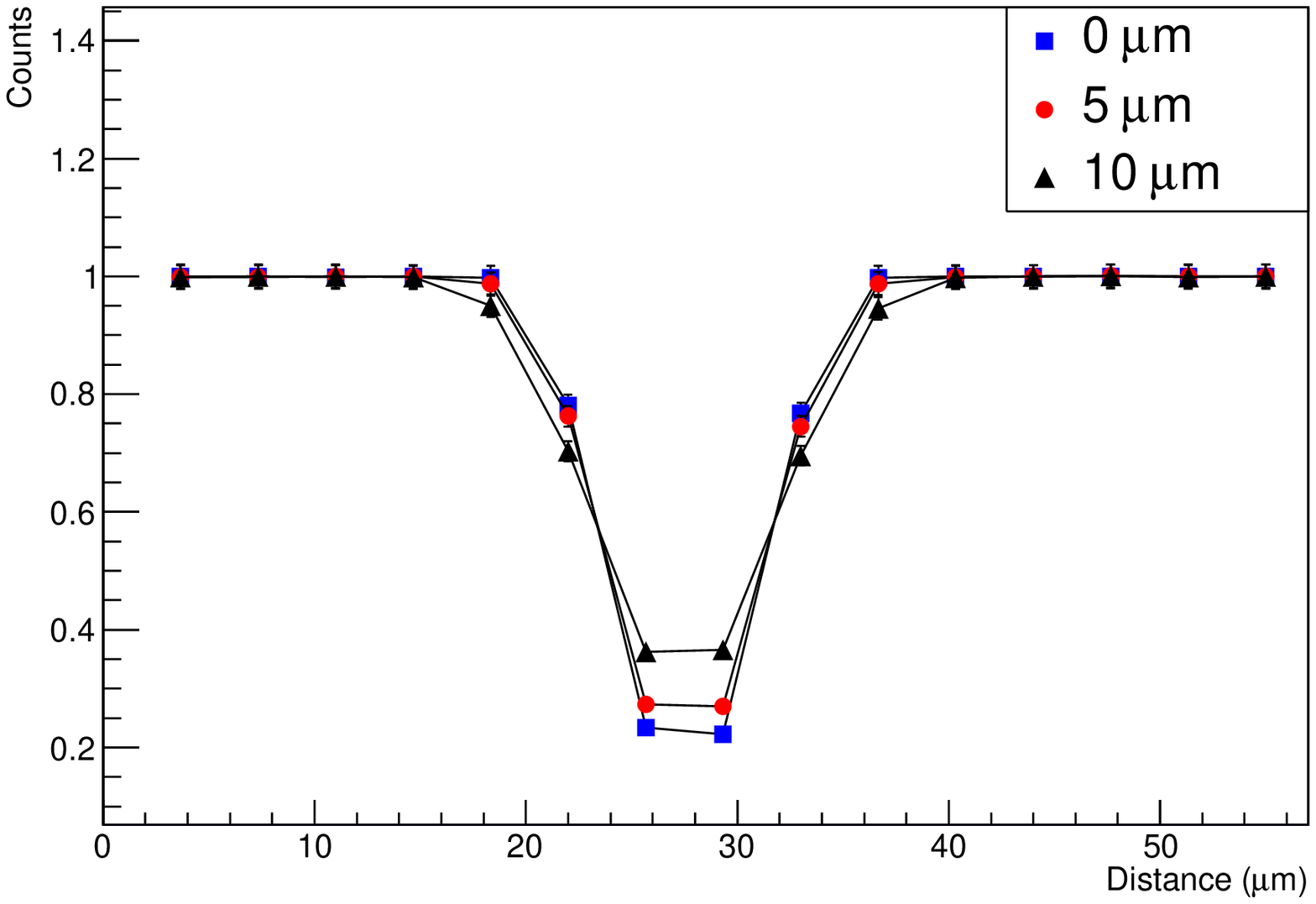}}
\end{center}
\caption{Number of counts as a function of beam position. Threshold values: 25\% (left), 50\% (center) and 75\% (right).} 
\label{chargecnt}
\end{figure}

Double counting due to charge sharing can occur when the threshold value adopted is low when compared to the deposited energy. However, as noticed from the Figure \ref{sensitivity} charge sharing can reduce count rates if high threshold values are adopted. Figure \ref{chargecnt} shows the normalized number of counts as a function the beam position. The curves behavior is similar to the observed in \citep{macraighne}. Three threshold values were used: 25\%, 50\% and 75\% with respect to the beam energy. The 25\% threshold curve (left) illustrates the excess due to multiple counts around 27.5 $\rm \mu m$. A very mild variation is seen for the 50\% threshold curve (center) as the amount of charge collected is shared by a factor of 1/2 by the 2 pixels at the midpoint which is exactly equal to the minimum necessary for a single pixel to be fired. The 75\% threshold curve (right) shows a decrease for counts in the same region.

Note that charge collection efficiency was considered constant along the whole sensor in the simulation. Although this is not necessarily correct in regions between pixels for certain device setups, it can be accounted in the simulation by simply modifying the data acquisition module description. Local threshold asymmetries between neighbor pixels can also be included to reproduce specific count profiles. 

The simulation also indicates that the size of the beam should be smaller than $\rm 5 \mu m$ such that no further de-convolutions are necessary for spatial resolution studies. This can also be inferred from equation \ref{diffusion} as the size of the beam becomes comparable with the size of charge carrier cloud at collection region.

\subsection{Modulation Transfer Function}

\begin{figure}[!htb]
\begin{center}
{\includegraphics[width=0.5\textwidth]{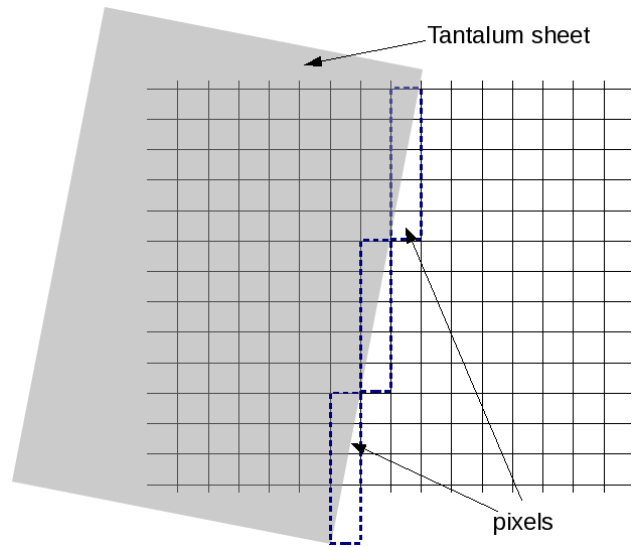}}
\end{center}
\caption{Pixel sensor surface partially covered by a thin Tantalum layer for MTF estimation.} 
\label{mtfsetup}
\end{figure}

A typical measurement to estimate the imaging performance of a pixel sensor is the determination of the modulation transfer function (MTF) which describes the spatial frequency response of the device. One of the methods to perform such experiment is by covering part of the sensor surface with a square sheet of a certain dense material (Tantalum or Tungsten) while the other part of the surface is uncovered. This piece of metal is placed with a small tilt with respect to the sensor plane as shown in Figure \ref{mtfsetup}. For the simulation, a thin sheet of Tantalum was placed in order to cover about a half of the sensor. Such high-Z material is used because of its low scattering probability. 

\begin{figure}[!htb]
\begin{center}
{\includegraphics[width=0.5\textwidth]{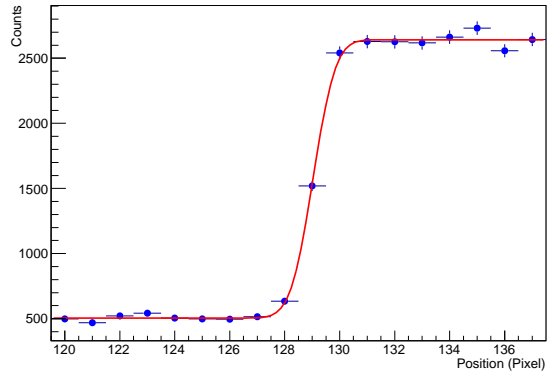}}
\end{center}
\caption{Pixel count profile obtained with the sensor partially covered by a Tantalum sheet.} 
\label{mtfedge}
\end{figure}

The whole sensor was irradiated in the simulation and the accumulated hit count per pixel was written in a 256 x 256 2D histogram. The integrated number of counts was obtained for each group of lines with the edge of the sheet dividing the pixels in the same column. Each group of lines was then aligned to the same column by shifting the indexes and an overall pixel count profile was obtained as shown in Figure \ref{mtfedge} where the red line illustrates the fit obtained for equation \ref{edge} used to model the count profile. 

\begin{figure}[!htb] 
\begin{center}
{\includegraphics[width=0.5\textwidth]{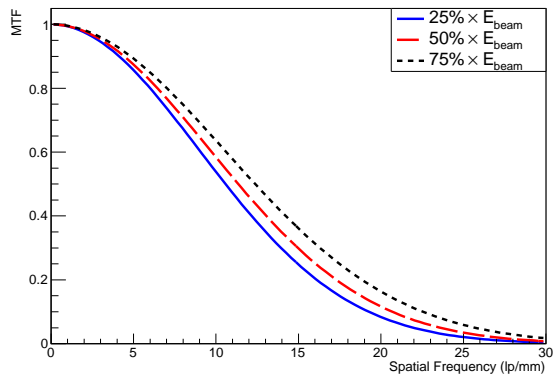}}
\end{center}
\caption{MTF for three different threshold values adopted. } 
\label{mtf}
\end{figure}

\begin{equation}\label{edge}
N(x) = A+B~{\rm erf}\left({(x-\mu)/\sqrt{2}\sigma}\right),
\end{equation}
where $A$ is the average count per pixel in the covered area, $B$ is the difference between the average counts for the uncovered and covered areas, $\mu$ is the center of the $\rm erf$ function and $\sigma$ determines the slope of the edge. The edge spread function was obtained by taking the derivative of the fit and the MTF is calculated as its Fourier transform. Figure \ref{mtf} shows the MTF curve for the different threshold values used (25\%, 50\%, 75\% of $\rm E_{beam}$). The spatial resolution defined as the spatial frequency at 70\% of the normalised MTF was $\rm 7.6\pm 0.5 ~lp/mm$, $\rm 8.1 \pm 0.7 ~lp/mm$ and $\rm 8.9 \pm 1.1 ~lp/mm$, respectively. The curves obtained are very similar to the experimentaly obtained at \citep{MTF1}.

As seen in section \ref{pixelsens}, the increase of the threshold causes a reduction of the effective active area of the pixels and as a consequence the spatial resolution improves accordingly.

Recently the same type of measurements were performed with CdTe hybrids to evaluate how parameters such as applied bias affect the spatial resolution of the device \citep{dima}. Although the shape of the MTF is similar and no significant back scattering is observed, the overall value for the spatial resolution is smaller for these devices as a consequence of the need for better energy resolution with thicker sensors. 

\section{Conclusions}\label{conclusion}

Despite the simple approach proposed in this work it reproduces many aspects of a variety of measurements. Results shown were in reasonable agreement with measurements and the simulation also provided additional information concerning interaction effect components in the material and reconstruction features of the studied device provided Monte Carlo truth was made available. The observed robustness of the method indicates that it can be used as a versatile and reliable tool for future experiments and design of similar devices. In particular, similarities observed between silicon and high-Z sensor measurements indicate that the same simulation could be adopted for these types of sensors. Energy measurement capabilities are also an important feature that is reproduced in the simulation and can be explored together with charge sharing characterisation.

\section*{Acknowledgments}

The authors would like to thank FAPERJ and CNPq agencies for financial support.

\bibliographystyle{elsarticle-num}
\bibliography{si_det_paper.bib}

\end{document}